\documentclass[preprint,aps,draft,linenumbers,showpacs,aps]{revtex4}

\usepackage{amsfonts}
\usepackage{amsmath}
\usepackage{bm}
\usepackage{graphicx}  
\usepackage{color}

\newcommand\Bv{{\bf B}}
\newcommand\Ev{{\bf E}}

\newcommand\uv{{\bf u}}
\newcommand\vv{{\bf v}}

\newcommand\jv{{\bf j}}

\begin{document}


\title{Vlasov simulations of Kinetic Alfv\'en Waves at proton kinetic scales}

\author{C. L. V\'asconez$^{1,2}$, F. Valentini$^1$, E. Camporeale$^3$ and P. Veltri$^1$}

\affiliation{
$^1$Dipartimento di Fisica, Universit\`a della Calabria, I-87036 Cosenza, Italy\\
$^2$Observatorio Astron\'omico de Quito, Escuela Polit\'ecnica Nacional, Quito, Ecuador\\
$^3$Centrum Wiskunde \& Informatica, Amsterdam, Netherlands
} 
\date{\today} 
\input epsf 
\begin{abstract} 
Kinetic Alfv\'en waves represent an important subject in space plasma physics, since they are thought to play a crucial role in the development of the turbulent energy cascade in the
solar wind plasma at short wavelengths (of the order of the proton inertial length $d_p$ and beyond). A full understanding of the physical mechanisms which govern the kinetic plasma dynamics at these
scales can provide important clues on the problem of the turbulent dissipation and heating in collisionless systems. In this paper, hybrid Vlasov-Maxwell simulations are employed to analyze in detail
the features of the kinetic Alfv\'en waves at proton kinetic scales, in typical conditions of the solar wind environment. In particular, linear and nonlinear regimes of propagation of these
fluctuations have been investigated in a single-wave situation, focusing on the physical processes of collisionless Landau damping and wave-particle resonant interaction. Interestingly,
since for wavelengths close to $d_p$ and proton plasma beta $\beta$ of order unity the kinetic Alfv\'en waves have small phase speed compared to the proton thermal velocity, wave-particle interaction
processes produce significant deformations in the core of the particle velocity distribution, appearing as phase space vortices and resulting in flat-top velocity profiles. Moreover, as the Eulerian
hybrid Vlasov-Maxwell algorithm allows for a clean almost noise-free description of the velocity space, three-dimensional plots of the proton velocity distribution help to emphasize how the plasma
departs from the Maxwellian configuration of thermodynamic equilibrium due to nonlinear kinetic effects. 
\end{abstract}
\pacs{?}
\maketitle

\section{Introduction}

The solar wind is a turbulent plasma \cite{marschcarb}, mainly composed by protons and electrons, which can be considered collisionless in good approximation. Because of the turbulent character of
the interplanetary medium, the energy injected into it at large scales as Alfv\'enic fluctuations is transferred towards short scales along the turbulent spectrum. In such a collisionless medium, what
physical mechanism drives the short-scale dissipation of the energy injected at large scales still remains an unanswered question and attracts nowadays a significant scientific interest. In fact, the
identification of the fluctuations responsible for channeling the energy toward short wavelengths and the full understanding of the dissipation mechanisms in the solar wind represent two top priority
subjects in space plasma physics.

The power spectrum of the solar-wind fluctuating fields in the range of long wavelengths  manifests a behavior reminiscent of the $k^{-5/3}$ Kolmogorov power law for fluids
\cite{kolmogorov41,coleman68,dobrowolny80,tu95,goldstein95}, $k$ being the wavenumber. The Kolmogorov-like spectral behavior extends down to a range of wavelengths close to typical proton kinetic
scales (the proton inertial length $d_p$, and/or the proton Larmor radius). Here, the features of the spectra abruptly change with the appearance of a spectral break
\cite{leamon00,bourouaine12,bale05} and kinetic effects presumably govern the system dynamics.

In these range of scales and even down to typical electron kinetic scales, many solar-wind observational analyses \cite{bale05,sahraoui09,podesta12,salem12,chen13,kiyani13}, theoretical works
\cite{howes08-1,schekochihin09,sahraoui12} and numerical simulations \cite{gary04,howes08-2,tenbarge12} suggest that the so-called Kinetic Alfv\'en waves (KAWs) can play an important role in the
mechanism of turbulent energy dissipation and heating. KAWs are Alfv\'en cyclotron waves propagating almost perpendicularly to the background magnetic field. An extensive linear analysis of these
waves has been performed by Hollweg in 1999 \cite{hollweg99} (see also references therein for a more complete view on the subject). The fact that these fluctuations can be important in the
development of the solar-wind turbulent cascade is supported by observational data which show that the distribution of wavevectors of long wavelength magnetic fluctuations has a significant
population
quasi-perpendicular to the ambient magnetic field \cite{matthaeus86,matthaeus90}.

In this paper we make use of the hybrid Vlasov-Maxwell (HVM) code \cite{valentini07}, to study numerically the characteristics of the KAWs in linear and nonlinear regime, in the range of spatial
scales close to proton skin depth. The HVM algorithm integrates numerically the Vlasov equation for the proton distribution function in multi-dimensional phase space. In the present work, we restrict
our analysis to the 1D-3V (one dimension in physical space and three dimensions in velocity space) phase space configuration. Within the HVM model the electrons are considered as a fluid and a
generalized Ohm equation is employed for computing the electric field, which retains the Hall term and the electron inertia effects. Quasi neutrality is assumed and the displacement current is
neglected in the Ampere equation, making the assumption of low frequency dynamics. Finally, an isothermal equation of state for the scalar electron pressure is employed to close the HVM system (for
more details on the HVM equations and on the numerical algorithm see Ref.\cite{valentini05-1,valentini07}). The HVM code is a well-tested algorithm which has been successfully employed for numerical
studies of plasma turbulence \cite{valentini10,servidio12,servidio14,matthaeus14,valentini14,greco12,perrone13} and ion-cyclotron heating \cite{iazzolino10}.

We present the results of a series of linear and nonlinear simulations, in a situation of single-wave propagation, with a single wavenumber slightly larger than the proton skin depth wavenumber.
Our goal is to perform a detailed analysis of the process of wave-particle interaction, in linear and nonlinear regime, involving resonant protons and KAWs, in typical conditions of the solar wind
environment and in a range of spatial scales close to $d_p$. As discussed in detail in the following, our numerical results show that the efficient resonant interaction between the KAWs and the
protons can significantly shape the particle distribution function, making the plasma depart from the Maxwellian configuration of thermodynamical equilibrium. 

This paper is organized as follows. In Section II the dispersion relation of the KAW and corresponding eigenmodes for magnetic and velocity perturbations are derived in the framework of linear
two-fluid theory, under the assumptions of quasi-neutrality and negligible displacement current. Section III is devoted to the description of the hybrid Vlasov-Maxwell simulations and of the numerical
results for the propagation of the KAWs, both in linear and nonlinear regimes. We conclude and summarize in Section IV.

\section{Two-fluid dispersion relation and eigenmodes}\label{twofluid}
In this Section, we revisit the two-fluid approach to derive the dispersion relation and the eigenmodes of the KAWs in linear approximation. To make contact with the HVM model, the set of linear
two-fluid (protons and electrons) equations have been solved under the assumption of quasi-neutrality ($n_e\simeq n_p=n$) and by neglecting the displacement current in the Ampere equation. Moreover,
proton and electron scalar pressures have been assigned a general adiabatic equation of state. This analysis allows us to specialize to the hybrid case the linear expectations for the wave frequency
of KAWs obtained in previous works (see, for instance, Ref. \cite{hollweg99}). The linear results obtained in this Section will be used to guide the HVM simulations discussed in detail in the
following. In particular, the linear frequency and the expressions of the eignemodes for the KAW branch will be employed to initialize the HVM simulations both in linear and nonlinear regime.

\begin{figure} 
\epsfxsize=7.5cm
\centerline{\epsffile{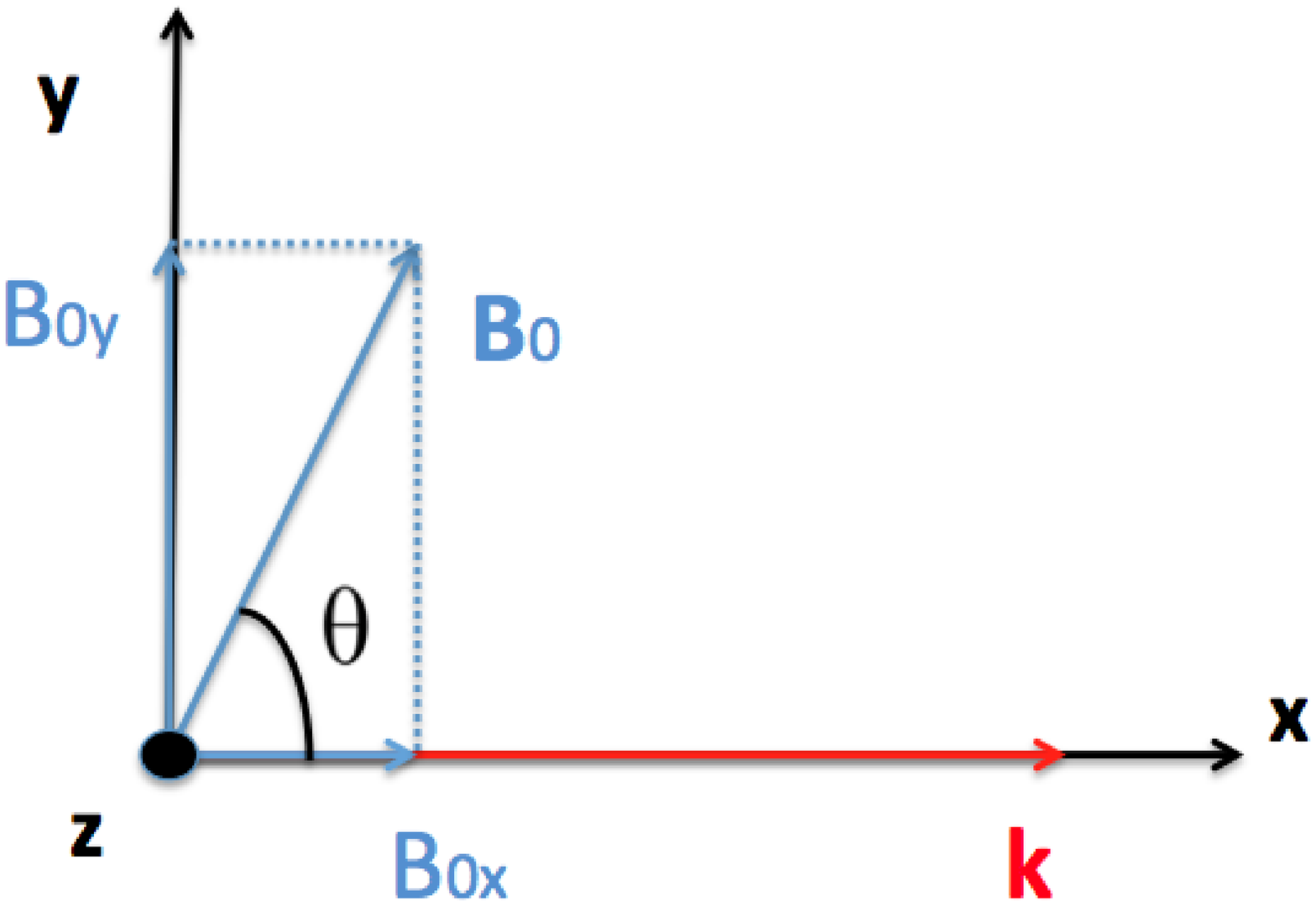}} 
\caption{(Color online) The reference frame chosen for the study of KAWs; the wavevector ${\bf k}$ is along the $x$ direction, while the background magnetic field ${\bf B}_0$ lies in the $x$-$y$
plane, inclined at
an angle $\theta$ with respect to ${\bf k}$.}
\label{fig1} 
\end{figure}

We choose the reference frame pictured in Fig. \ref{fig1}, in which the wave vector ${\bf k}=(k,0,0)$ is along the positive $x$ direction and the background magnetic field ${\bf
B}_0=(B_{0x},B_{0y},0)$ lies in the $x$-$y$ plane, inclined at an angle $\theta$ with respect to the $x$ axis. In these conditions the problem is intrinsically one dimensional in physical space.

By coupling the continuity equations for particle density and momentum equations to Maxwell equations for fields, under the assumptions discussed above, after some algebra, one can obtain the
dimensionless wave dispersion relation in the form of a sixth-order polynomial equation, which can be written as:

\begin{equation}
A\omega^6+B\omega^4+C\omega^2+D=0
\label{reldisp}
\end{equation}

where:

\begin{eqnarray*}
A=& &(1+k^2d_e^2)^2 \\\nonumber
B=& &-(1+k^2d_e^2)k^2[1+\cos^2{\theta}+\beta (1+k^2d_e^2)]+\\
& &-k^4\cos^2{\theta}\\
C=& &k^4\cos^2{\theta}[1+\beta k^2+2(1+k^2d_e^2)]\\
D=& &\beta k^6\cos^4{\theta}
\end{eqnarray*}

The above equations have been rescaled by normalizing time by the inverse proton cyclotron frequency $\Omega_{cp}^{-1}$, velocities by the Alfv\'en speed $V_{_A}$, mass by the proton mass $m_p$, and
lengths by the proton inertial length $d_p=V_{_A}/\Omega_{cp}$. In these units, the electron inertial length is given by $d_e=(m_e/m_p)^{1/2}$. Also, the modulus of the background magnetic field is
set $B_0=1$. For now on, all physical quantities will be rescaled by the characteristic parameters listed previously.

\begin{figure} 
\epsfxsize=8cm
\centerline{\epsffile{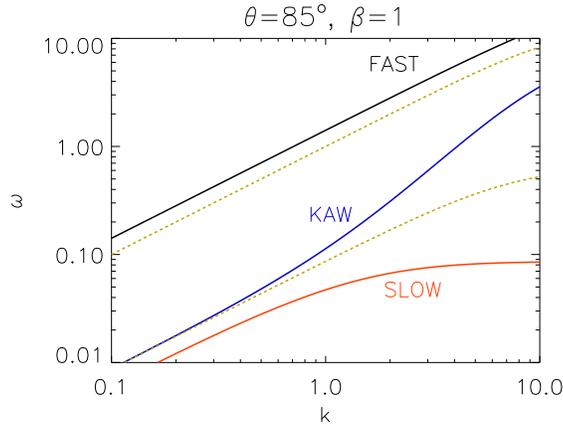}} 
\caption{(Color online) The three roots of Eq. (\ref{reldisp}), i. e. FAST (black-solid curve), SLOW (red-solid curve) and KAW (blue-solid curve) branches, for $\theta=85^\circ$, $\beta=1$ and in the
range of
wavenumbers around $k=1$. The green-dashed curves represent the solutions of Eq. (\ref{reldisp}), in the case of cold plasma ($\beta\ll 1$), for which the SLOW branch disappears.}
\label{fig2} 
\end{figure}

Eq. (\ref{reldisp}) can be solved analytically by making use of the so-called Vieta's substitution \cite{vieta}, to obtain values of the wave frequencies for any given angle
$\theta$, any value of $\beta$ and for any range of wave numbers. In Fig. \ref{fig2}, we show, in a logarithmic plot, the three roots of Eq. (\ref{reldisp}), obtained for positive values of
$\omega$, for a specific case with $\theta=85^\circ$ and $\beta=1$, in the range of wavenumbers around the proton skin depth wavenumber ($k=1$). In this plot, the black (red) solid line represents
the branch of fast (slow) magnetosonic waves, while the blue solid line refers to the KAW branch. The green-dashed curves are the solutions obtained under the cold plasma approximation ($\beta\ll
1$), for which the slow magnetosonic branch disappears.

Moreover, from the analysis of the linearized two-fluid equations, one can also get explicit expressions of the eigenmodes for magnetic ($\delta {\bf B}$) and velocity ($\delta {\bf u}$)
perturbations, which read:

\begin{eqnarray}\label{eigen1}
\delta B_z&=& 2a\cos{(kx)}\\
\delta B_y&=&-\frac{\cos{\theta}}{\omega}\left [1-\frac{\omega^2(1+k^2d_e^2)}{k^2\cos^2{\theta}}\right ]2a\sin{(kx)}\\
\delta u_z&=&-\frac{k\cos{\theta}}{\omega}2a\cos{(kx)}\\
\delta u_y&=& \frac{k\cos^2{\theta}}{\omega^2}\left [1-\frac{\omega^2(1+k^2d_e^2)}{k^2\cos^2{\theta}}\right ]2a\sin{(kx)}\\\nonumber
\delta u_x&=& \frac{k\sin{(2\theta)}}{2\left(\omega^2-\beta k^2\right)}\left [1-\frac{\omega^2(1+k^2d_e^2)}{k^2\cos^2{\theta}}\right ]\times \\
& &2a\sin{(kx)}
\label{eigen5}
\end{eqnarray}

where $a$ is real number.

It is worth noting from the expressions above that the magnetic and the velocity eigenmodes have elliptic polarization and a $\pi/2$ phase displacement.

The previous expressions for magnetic and velocity perturbations will be used to initialize the HVM simulations of KAWs, presented in the next Section. Clearly, Eqs. (\ref{eigen1})-(\ref{eigen5})
are not exact eigenmodes of the HVM equations; nevertheless, as we will show in the following, when used as initial perturbations in the HVM simulations, they allow to excite predominantly one
desired wave mode, by selecting the appropriate value of $\omega$ for a given $k$. For our purpose of studying numerically the features of the KAWs, this is
crucial since it prevents the excitation of mixture of modes which would complicate the analysis.

\section{Hybrid Vlasov-Maxwell simulations}
We solve numerically the HVM equations \cite{valentini07} in 1D-3V phase space configuration. The set of HVM equations in dimensionless units can be summarized in the form:

\begin{eqnarray}
  \label{eq:vlasov}
& & \frac{\partial f}{\partial t} + \vv \cdot {\nabla} f+ \left( \Ev + \vv \times \Bv \right)\cdot\frac{\partial f}{\partial \vv} = 0\\\nonumber
& &{\bf E} - d_e^2\Delta {\bf E}  = - ({\bf u} \times {\bf B}) + \frac{1}{n}({\bf j} \times {\bf B}) -\frac 1 n \nabla P_e +\\
\label{eq:ohm3}
&+& \frac{d_e^2}{n}\left[\nabla\cdot {\bf \Pi}+\nabla\cdot({\bf u}{\bf j}+{\bf j}{\bf u})-\nabla \cdot \left( \frac{{\bf j}{\bf j}}{n} \right)\right]\\
\label{eq:Maxw_b}
& &\frac{\partial {\Bv}}{\partial t} = - \nabla \times {\Ev};\;\;\;\nabla \times {\Bv} = {\jv}
\end{eqnarray}

\noindent
where $f$ is the proton distribution function, $\Ev$ and $\Bv$ the electric and magnetic fields, respectively and ${\bf j}$ the total current density
(the displacement current has been neglected in the Ampere equation and quasi-neutrality is assumed). 
In Eq. (\ref{eq:ohm3}) the following compact notations have been used (see Ref. \cite{valentini07} for more details):
\begin{eqnarray}
& & [\nabla\cdot({\bf u}{\bf j}+{\bf j}{\bf u})]_i = \frac{\partial }{\partial x_j} (u_i j_j + j_i u_j ) \\
& & \left [\nabla \cdot \left( \frac{{\bf j}{\bf j}}{n} \right)\right]_i= \frac{\partial }{\partial x_j} \left(\frac{j_i j_j }{n}\right)
\end{eqnarray}
In 1D-3V configuration, the spatial variations occur along the $x$-direction, but each vector has three components; in this case, $\nabla=d/dx$, $\Delta=d^2/dx^2$.
The proton density $n$, bulk velocity ${\uv}$ and pressure tensor ${\bf\Pi}$ are obtained as velocity moments of $f$. The scalar electron pressure $P_e$ is assigned an isothermal equation of state
$P_e = n T_e$, where $T_e$ is the electron temperature. As in the previous Section, the background magnetic field is chosen to lie in the $x$-$y$ plane (see Fig. \ref{fig1}). 

The numerical algorithm employed to solve the above HVM equations (\ref{eq:vlasov})-(\ref{eq:Maxw_b}) is based on the coupling of the well-known splitting method \cite{cheng76} and the Current Advance
Method \cite{matthews94} for the electromagnetic fields, generalized to the hybrid case in Ref.\cite{valentini07}. Periodic boundary conditions are employed in physical space, while in the
velocity domain the distribution function is set equal to zero at $|v|>v_{max}$, where $v_{max}$ fixes the limits of the numerical domain in each velocity direction. For each simulation discussed in
the following, the time step $\Delta t$ has been chosen in such a way to satisfy the Courant-Friedrichs-Levy condition \cite{peyret83}, for the numerical stability of time explicit finite difference
algorithms.

\subsection{Linear regime}
The numerical analysis of the linear regime of wave propagation, presented in this Section, is preparatory for the study of the nonlinear regime discussed in the following. These preliminary
simulations can be, therefore, considered as a benchmark to show that the HVM code is able to describe properly the evolution of the KAWs and that the numerical resolution employed is
adequate to ensure a satisfactory conservation of the HVM invariants (energy, mass, entropy).

For the analysis of the linear regime, we simulate a plasma embedded in a uniform magnetic field ${\bf B}_0=B_{0x}{\bf e}_x+B_{0y}{\bf e}_y$ (see Fig. \ref{fig1}). At t=0, protons have
homogeneous and constant density and Maxwellian distribution of velocities. We set $\beta=2 v_{thp}^2/V_{_A}^2=1$ ($v_{thp}$ being the proton thermal speed), while the electron to proton temperature
ratio is $T_e/T_p=1$ and the proton to electron mass ratio is set $m_p/m_e=100$ (we point out that, for a realistic mass ratio, the electron skin depth $d_e$ cannot be adequately resolved with the
numerical resolution chosen for the HVM simulations). This initial condition is perturbed at $t=0$ by imposing on the system the magnetic and velocity perturbations in
Eqs. (\ref{eigen1})-(\ref{eigen5}), calculated for the KAW solution, with $a=10^{-5}$ and $k=mk_0$, where $k_0=2\pi/L$ is the fundamental wave number; we fixed $m=6$, thus $k=3d_p^{-1}$.

The length of the spatial box is $L=4\pi$, while $v_{max}=4.5v_{thp}$ in each velocity direction. The numerical 1D-3V phase space domain is discretized by $N_x=512$ grid points in physical space and
$N_{V_y}=N_{V_z}=41$ grid points along the $v_y$ and $v_z$ directions. The number of grid points used to discretized the $v_x$ direction has been chosen in such a way to avoid effects of numerical
recurrence. In fact, in the case of very weak electric and magnetic fields, as it happens in linear regime, Eq. (\ref{eq:vlasov}) describes a motion close to free streaming and its solution, in 1D
physical space, can be written as $f_k(v,t)=f_0(v)\exp{[ik(x-vt)]}$. If the mesh spacing in the $v_x$ velocity direction is $\Delta v_x=2v_{max}/N_{V_x}$, there is a numerical recurrence occurring at
$T_R=2\pi/(k\Delta v_x)$. Therefore, if $t_{max}$ is the maximum time of the simulation, $\Delta v_x$ must be chosen in such a way that $t_{max}<T_R$, for a fixed value of $k$. For example, for
$t_{max}=125$, $k=3$ and $v_{max}=4.5v_{thp}$, one must choose $N_{V_x}=401$ in such a way to have $T_R\simeq 132 > t_{max}$.

\begin{figure} 
\epsfxsize=8cm
\centerline{\epsffile{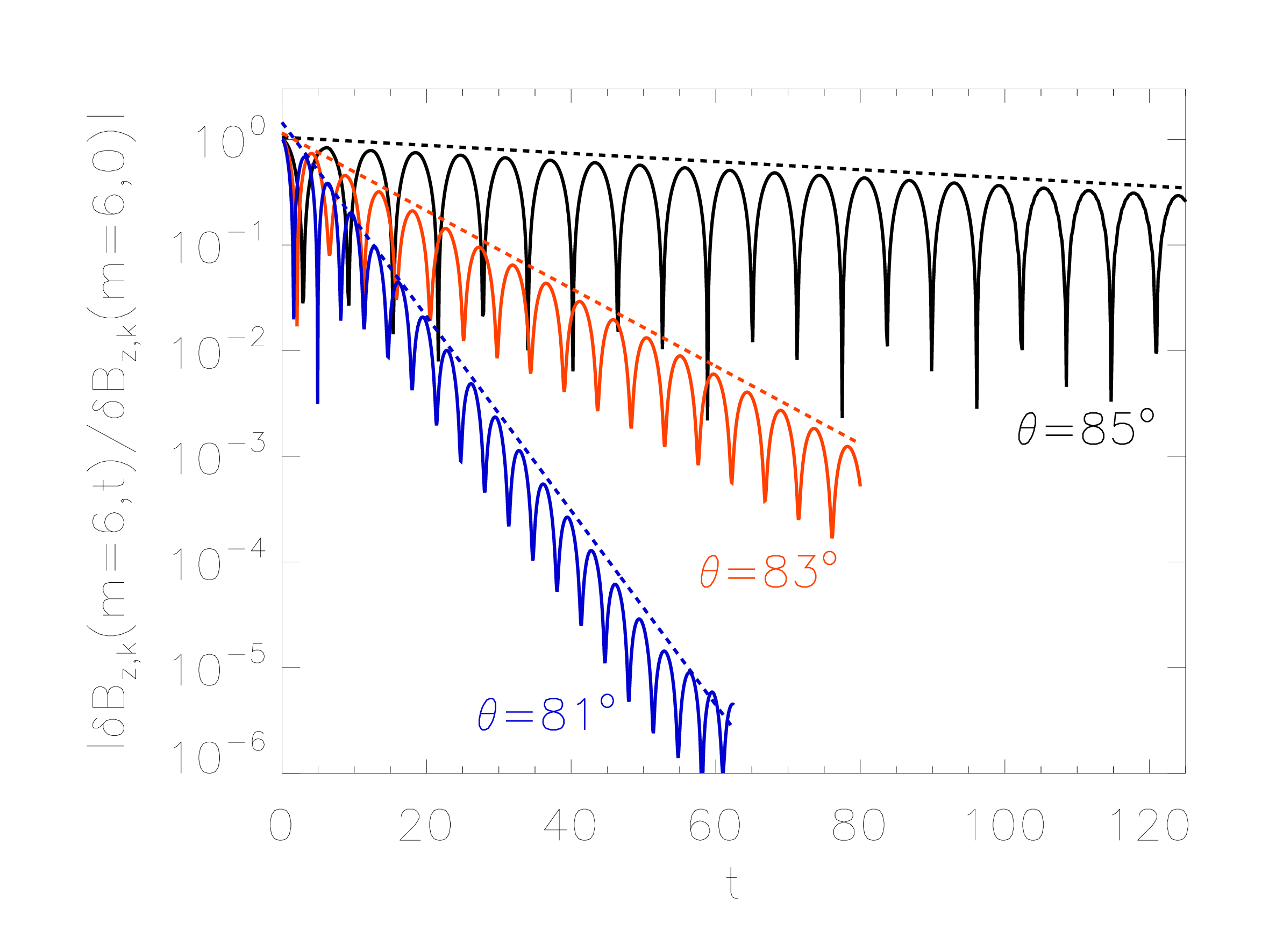}} 
\caption{(Color online) Time evolution of $|\delta B_{z,k}(m=6,t)/\delta B_{z,k}(m=6,0)|$, for $a=10^{-5}$ and for $\theta=81^\circ, 83^\circ, 85^\circ$ (blue-solid, red-solid and black-solid line,
respectively). The blue-dashed, red-dashed and black-dashed lines represent the best fits for the damping rates of the oscillations.}
\label{fig3} 
\end{figure}

From linear kinetic plasma theory \cite{krall86}, small-amplitude waves in a uniformly magnetized plasma undergo Landau damping if they have a component of propagation along the background magnetic
field ${\bf B}_0$; only particle moving along ${\bf B}_0$ contribute to damping, because in a uniform magnetic field there is no net motion of particles across the field. In order to analyze the
effects of Landau damping \cite{landau46} on the KAW oscillations, as dependent on the propagation angle with respect to ${\bf B_0}$, we performed 14 simulations for different values of $\theta$ (the
angle between the wave vector and the background magnetic field) in the range $69^\circ \leq \theta\leq 85^\circ$. In Fig. \ref{fig3} we report, in semilogarithmic plot, the
time evolution of the absolute value of the $m=6$ Fourier component of the magnetic fluctuation $\delta B_{z,k}(m=6,t)$, normalized to its value at $t=0$, for $\theta=81^\circ, 83^\circ,
85^\circ$ (blue-solid, red-solid and black-solid line, respectively). In agreement with linear kinetic theory \cite{krall86}, the oscillation amplitude decays exponentially in time, the damping rate
$\gamma$ appearing strongly dependent on $\theta$ (the smaller $\theta$, the larger $\gamma$). Note that for the simulation with $\theta=85^\circ$ we fixed $t_{max}=125$, while simulations with
smaller $\theta$, for which heavily damped oscillations are recovered, have been stopped at earlier time. The blue-dashed, red-dashed and black-dashed lines in this figure represent the best fits for
the damping rates of the oscillations, whose absolute values result $|\gamma|\simeq 0.24, 0.085, 0.009$ for $\theta=81^\circ, 83^\circ, 85^\circ$, respectively.

\begin{figure} 
\epsfxsize=8cm
\centerline{\epsffile{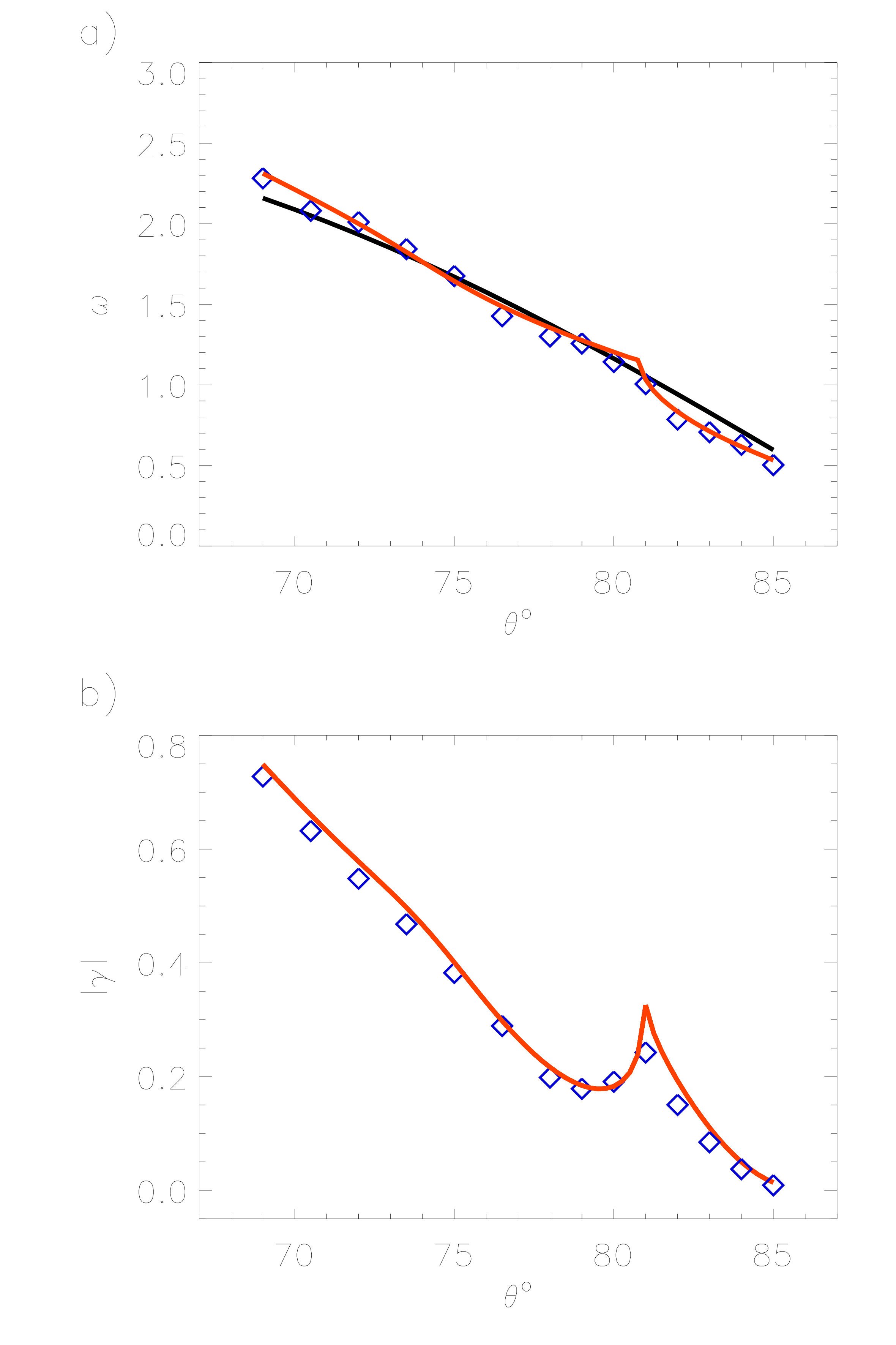}} 
\caption{(Color online) Panel a): dependence of the wave frequency $\omega$ on the propagation angle $\theta$, obtained from the linear HVM simulations (blue diamonds), linear two-fluid theory
(black-solid curve)
and linear kinetic Vlasov solver (red-solid curve). Panel b): dependence of the absolute value of the damping rate $|\gamma|$ on $\theta$, obtained from the linear HVM simulations (blue diamonds)
and from the linear kinetic Vlasov solver (red-solid curve).}
\label{fig4} 
\end{figure}

In order to show that the oscillations recovered in our simulations are in fact KAWs, in Fig. \ref{fig4}, we plot the numerical results for the oscillation frequency $\omega$ [blue diamonds
in panel a)], and for the absolute value of the damping rate $|\gamma|$ [blue diamonds in panel b)], as functions of $\theta$, for the 14 simulations performed. In panel a), we also reported the
theoretical prediction for $\omega=\omega(\theta)$ on the KAW branch obtained from the two-fluid approach (black-solid curve) discussed in Section \ref{twofluid} [see Eq. (\ref{reldisp})] and from a
kinetic linear Vlasov solver (red-solid curve). In panel b), the solution of $|\gamma|=|\gamma(\theta)|$ for the KAW branch, obtained from a kinetic linear Vlasov solver, is indicated by a
red-solid curve. We note that the linear Vlasov solutions have been obtained by employing the standard kinetic theory for an electron-proton plasma \cite{stix92}. However, in order to mimic the HMV
system, we have considered the limit $m_e/m_p \to 0$. Panels a) and b) in Fig. \ref{fig4} show a nice agreement between numerical results and analytical (two-fluid and kinetic) predictions,
demonstrating that the fluctuations recovered in our HVM simulations can be identified as KAWs.

To conclude this Section, we shortly discuss the conservation of the HVM invariants during the simulations. In particular, we consider the conservation of proton mass and entropy and total
energy. Concerning the latter quantity, we point out that for the HVM equations (\ref{eq:vlasov})-(\ref{eq:Maxw_b}), a standard conservation law cannot be derived for the total energy,
as a consequence of neglecting the displacement current in the Ampere equation. In fact, by multiplying the proton Vlasov Eq. (\ref{eq:vlasov}) by $v^2/2$, through integration over
the whole velocity space and over the 1D periodic spatial domain $D=[0,L]$ (the result can be easily generalized to the 3D case in physical space), one gets the following dimensionless equation:
\begin{equation}
\int_0^L \left[\frac{\partial}{\partial t}\left(\frac 3 2 n T_p +\frac 1 2 n u^2+\frac{B^2}{2}\right)+{\bf E}\cdot {\bf j}_e\right] dx=0
\label{energy1}
\end{equation}
where ${\bf j}_p=n{\bf u}$ and ${\bf j}_e=\nabla\times {\bf B}-{\bf j}_p$ are proton and electron current densities, respectively (we remind the reader that in scaled units the proton mass is $m_p=1$
and the proton electric charge is $e=1$). In Eq. (\ref{energy1}) one recognizes the contribution of the thermal energy density $3/2nT_p$, of the kinetic energy density $1/2nu^2$ and of
the magnetic energy density $B^2/2$, while the term ${\bf E}\cdot {\bf j}_e$ represents the work of the electric field on the electrons.

As anticipated before, Eq. (\ref{energy1}) is not in the usual form of conservation law; nevertheless, if one sets:
\begin{eqnarray}
& &L_e=\int_0^Ldx\int_0^t {\bf E}\cdot {\bf j}_e\; dt';\; E_{M}=\int_0^L\frac{B^2}{2}dx\\
& &E_{kin}=\int_0^L\frac{nu^2}{2}\; dx;\; E_{th}=\int_0^L\frac {3nT_p}{2}\; dx,
\end{eqnarray}
Eq. (\ref{energy1}) can be re-written in the following form:
\begin{equation}
\frac{\partial}{\partial t}(L_e+E_{th}+E_{kin}+E_{M})=0
\end{equation}
or, equivalently, as:
\begin{equation}
E_{tot}=L_e+E_{th}+E_{kin}+E_{M}=const.
\end{equation}
The quantities $L_e$, $E_{M}$, $E_{th}$ and $E_{kin}$ can be evaluated at each time step in the HVM simulations, in such a way to control the conservation of $E_{tot}$. 

For the linear simulations discussed in this Section, typical relative mass variations are limited to $\sim 10^{-5}\%$, entropy variations to $\sim 0.036\%$ and total energy variations to
$\sim 0.088\%$, confirming the adequacy of the numerical resolution adopted and the reliability of the numerical results.

\subsection{Nonlinear regime}

To investigate the nonlinear regime of propagation of the KAWs, we considered six different simulations, whose typical parameters ($\beta$, $T_e/T_p$, $m_p/m_e$), initial condition and initial
magnetic and velocity perturbations are the same as those described in the previous Section, except for the amplitude of the perturbations that is set now $a=0.15$, $0.17$, $0.19$, $0.21$,
$0.23$, $0.25$, respectively. The length of the spatial box is, as for the linear simulations, $L=4\pi$, while we set $v_{max}=5v_{thp}$. The number of grid points in the spatial domain is $N_x=512$
and in the three-dimensional velocity domain we set $N_{V_x}=N_{V_y}=N_{Vz}=91$. These nonlinear simulations follow the plasma dynamics up to a time $t_{max}=1000$. The typical relative variations of
mass, energy and entropy for these nonlinear simulations are $\sim 10^{-3}\%$, $\sim 1\%$ and $0.8\%$, respectively.

\begin{figure} 
\epsfxsize=8cm
\centerline{\epsffile{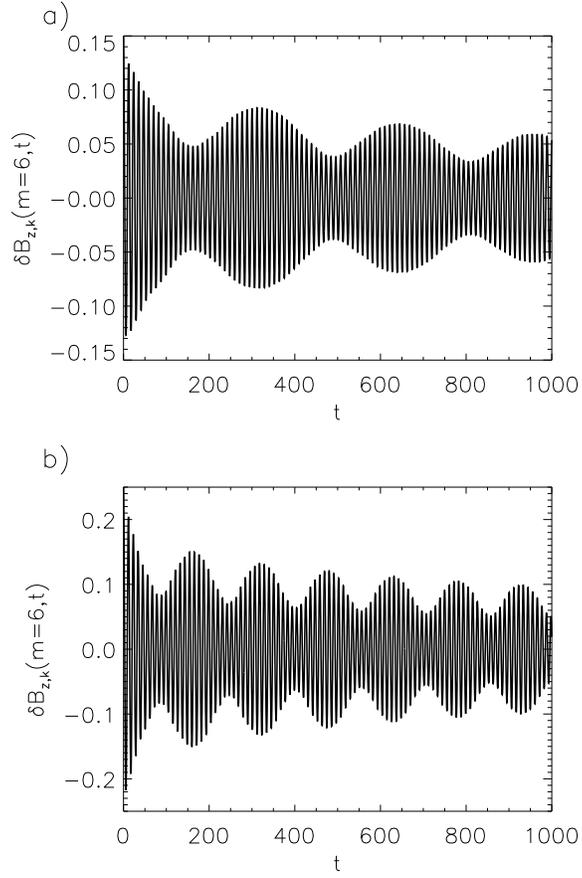}} 
\caption{Time evolution of $\delta B_{z,k}(m=6,t)$, for $a=0.15$ [panel a)] and $a=0.25$ [panel b)], and for $\theta=85^\circ$. }
\label{fig5} 
\end{figure}

In order to compare the evolution of the magnetic fluctuations for large initial perturbation amplitudes with that obtained in linear regime, in Fig. \ref{fig5} we plot $\delta B_{z,k}(m=6,t)$,
normalized to its value at $t=0$, for $\theta=85^\circ$, in the cases with $a=0.15$ [panel a)] and $a=0.25$ [panel b)]. In the linear case (see Fig. \ref{fig3}), the magnetic oscillations for
$\theta=85^\circ$ appear exponentially damped up to a time $t=120$. On the other hand, in the nonlinear regime pictured in Fig. \ref{fig5}, after a preliminary stage of exponential decay of the wave
amplitude, Landau damping is saturated by nonlinear effects and, for both $a=0.15$ and $a=0.25$, the magnetic fluctuations display characteristic envelope oscillations, whose period appears to be
inversely proportional to $a$ (larger values of $a$ correspond to smaller envelope oscillation periods). It is worth noting that no decay instability toward smaller wavenumbers has been recovered
during the simulation, even though the wavelength of the imposed perturbations is not the largest wavelength that fits in the spatial simulation box; the KAW mode excited at $t=0$ remains dominant
with a small amuont of energy stored in its harmonics.

\begin{figure} 
\epsfxsize=8cm
\centerline{\epsffile{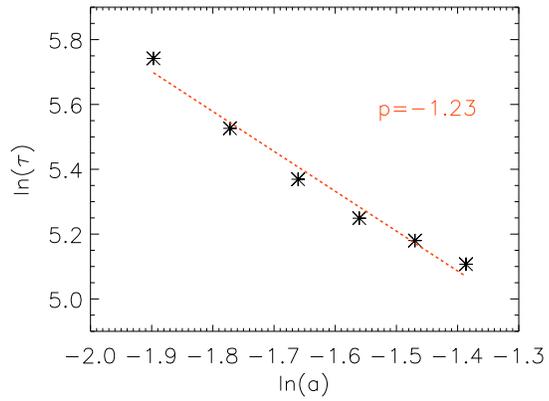}} 
\caption{(Color online) Dependence of the logarithm of the wave envelope oscillation period on the logarithm of the initial perturbation amplitude.}
\label{fig6} 
\end{figure}

The phenomenology described above is clearly reminiscent of the nonlinear saturation of Landau damping of electrostatic waves, due to particle trapping
\cite{oneil65,manfredi97,brunetti00,valentini05-2}. We analyzed in detail the dependence of the period $\tau$ of the envelope oscillations displayed in Fig. \ref{fig5} on the amplitude $a$ of the
initial perturbations. Assuming a relation of the type $\tau=a^p$, in Fig. \ref{fig6} we show $\ln{(\tau)}$ versus $\ln{(a)}$ for the six nonlinear simulations (black stars). The red-dashed line in
this figure represents the best fit of the numerical data; the best-fitting procedure gives a value $p=-1.23$.

Simple arguments help to understand how protons can be trapped by a pseudo-potential and give rise to envelope oscillations in the wave amplitude, as shown in Fig. \ref{fig5} and in analogy with the
electrostatic case. Assuming spatial variations only in the $x$ direction, in a reference frame moving with the wave phase speed $v_\phi$, the electric potential $\phi$ can be seen as a static
potential, depending only on a single variable $\xi=x-v_\phi t$. For a single proton with velocity ${\bf v}=(v_x,v_y,v_z)$, one can then derive a dimensionless energy conservation law, in the form:
\begin{eqnarray} \label{entrap}
 & &\mathcal{E}=\frac{(v_x-v_\phi)^2+v_y^2+v_z^2}{2}+\phi(\xi)=const.\\
 & &\phi(\xi)=-\int_{\xi_0}^{\xi} E_x(\xi ') d\xi '
\end{eqnarray}
$\xi_0$ being an arbitrary constant.

Moreover, conservation equations for the canonical momentum in $y$ and $z$ directions can be used:
\begin{eqnarray}\label{mom1}
 & &v_y+A_y(\xi)=v_{y0}; \; v_z+A_z(\xi)=v_{z0}\\
 & &A_y(\xi)=\int_{\xi_0}^{\xi}  B_z(\xi ') d\xi '\\ 
 & &A_z(\xi)=-\int_{\xi_0}^{\xi}  B_y(\xi ') d\xi '
\end{eqnarray}
$v_{y0}$ and $v_{z0}$ being two constants, and $A_y(\xi)$, $A_z(\xi)$ the $y$ and $z$ components of the magnetic potential. Eqs. (\ref{mom1}) allow to express $(v_y^2+v_z^2)/2$ as a function of
$\xi$ and, from Eq. (\ref{entrap}), one gets:
\begin{eqnarray}\nonumber
 \mathcal{E}(\xi,v_x)&=&\frac{(v_x-v_\phi)^2}{2}+\phi(\xi)+\frac 1 2 [v_{y0}-A_y(\xi)]^2+\\
 &+&\frac 1 2[v_{z0}-A_z(\xi)]^2=const.
\end{eqnarray}
The above equation can be re-written as:
\begin{equation}
 \mathcal{E}(\xi,v_x)=\frac{(v_x-v_\phi)^{2}}{2}+\Phi(\xi)=const.
\end{equation}
Here, $\Phi(\xi)=\phi(\xi)+[v_{y0}-A_y(\xi)]^2/2+[v_{z0}-A_z(\xi)]^2/2$ can be viewed as a pseudo-potential which can trap resonant protons. 

As the HVM code allows for a clean low-noise description of the proton distribution function, the role of nonlinear kinetic effects on the plasma dynamics, and in particular of the trapping of
protons in the pseudo-potential $\Phi$, can be directly observed in phase space contour plots or in three-dimensional velocity iso-surface plots of $f$. 

\begin{figure} 
\epsfxsize=8cm
\centerline{\epsffile{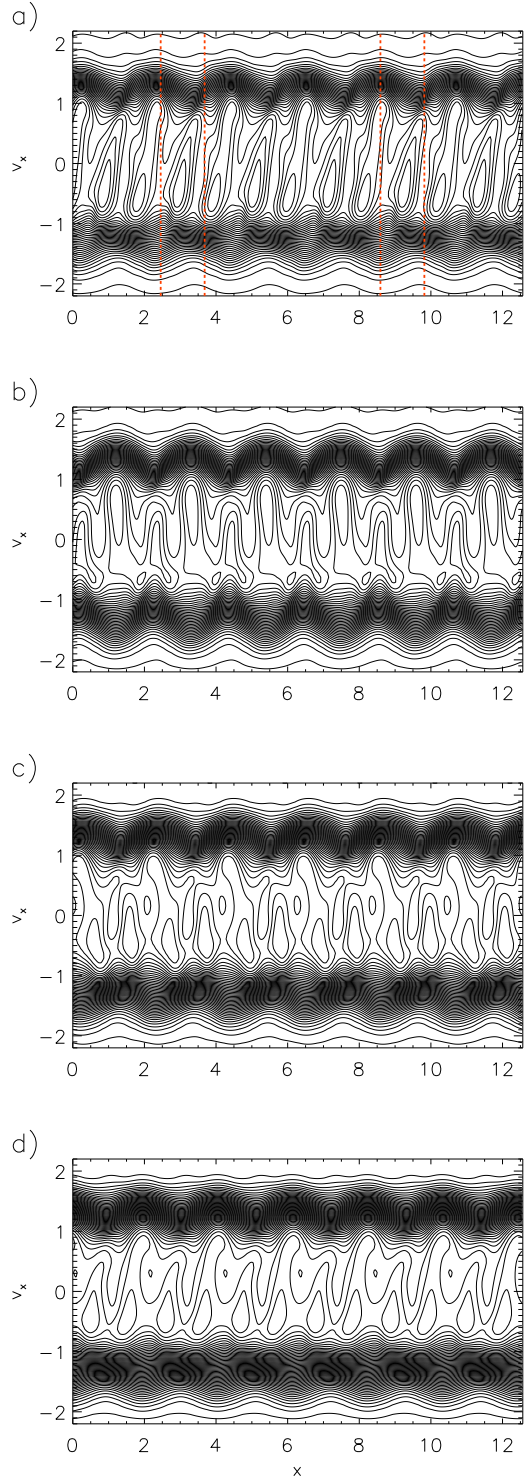}} 
\caption{(Color online) $x-v_x$ level lines of the proton distribution function $f$ calculated at $v_y=v_z=0$ at four different times $t=300, 500, 700, 1000$ (from top to bottom), for the nonlinear
simulation with
$a=0.25$.}
\label{fig7} 
\end{figure}

In Fig. \ref{fig7} we report the $x-v_x$ level lines of $f$ calculated at $v_y=v_z=0$ (i. e., we selected the phase space along the direction of ${\bf k}$) at four different times for the simulation with
$a=0.25$ ($t=300, 500, 700, 1000$ from top to bottom). The phase velocity of the fluctuations, evaluated through the Fourier analysis on the numerical signals, is $\omega/k\simeq 0.2$. In panel a) of
Fig. \ref{fig7}, vortical phase space structures are clearly visible in the velocity range around $v_x=0$. These phase space vortices are typical signature of the presence of trapped particle
populations \cite{manfredi97,brunetti00,valentini05-2}. The resulting complicated phase space contour lines of $f$ are determined by the nonlinear interaction and beating of two counter-propagating
signals (in our initial condition both positive and negative values of the wavenumber are excited for a given positive frequency). At larger times, in panel b), c), and d), the phase space structures
with positive mean velocity moves along the positive $x$ direction, while the one with negative mean velocity moves in the negative $x$ direction, giving rise to different phase space shapes. Due to
the fact that the phase velocity of the fluctuations is small compared to both $V_{_A}$ and $v_{thp}$, these strong phase space distortions are located in the middle of the core of the proton
distribution function, confined in the velocity range $-1\lesssim v_x\lesssim 1$ for each spatial position.

\begin{figure} 
\epsfxsize=8cm
\centerline{\epsffile{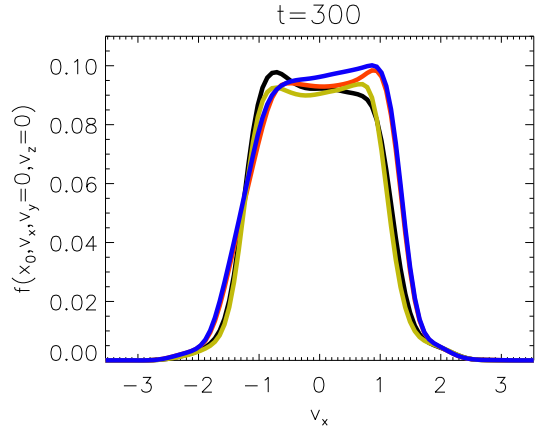}} 
\caption{(Color online) $v_x$ profiles of the proton distribution function $f$ calculated at $v_y=v_z=0$ and at four different spatial positions $x_0\simeq 2.5, 3.5, 8.5, 10$ (red, black, blue and
green, respectively) for the nonlinear simulation with $a=0.25$, at $t=300$.}
\label{fig8} 
\end{figure}

Figure \ref{fig8} shows four different slices of the $x-v_x$ proton distribution function at $t=300$, evaluated at four different spatial positions $x_0\simeq 2.5, 3.5, 8.5, 10$ (red, black, blue and
green, respectively), indicated by vertical red-dashed lines in Fig. \ref{fig7} a). It is clear from the curves in Fig. \ref{fig8} that the nonlinear wave-particle interaction, occurred in the
velocity range $-1\lesssim v_x\lesssim 1$,  has produced peculiar flat-top velocity profiles, by flattening the peak of the $v_x$ proton velocity distribution. It is worth to point out that the
velocity width of the flat-top region is nearly independent on $x$, as it can be deduced from Fig. \ref{fig8}.

\begin{figure*}[htpb]
$\begin{array}{ccc}
\epsfxsize=7cm \epsffile{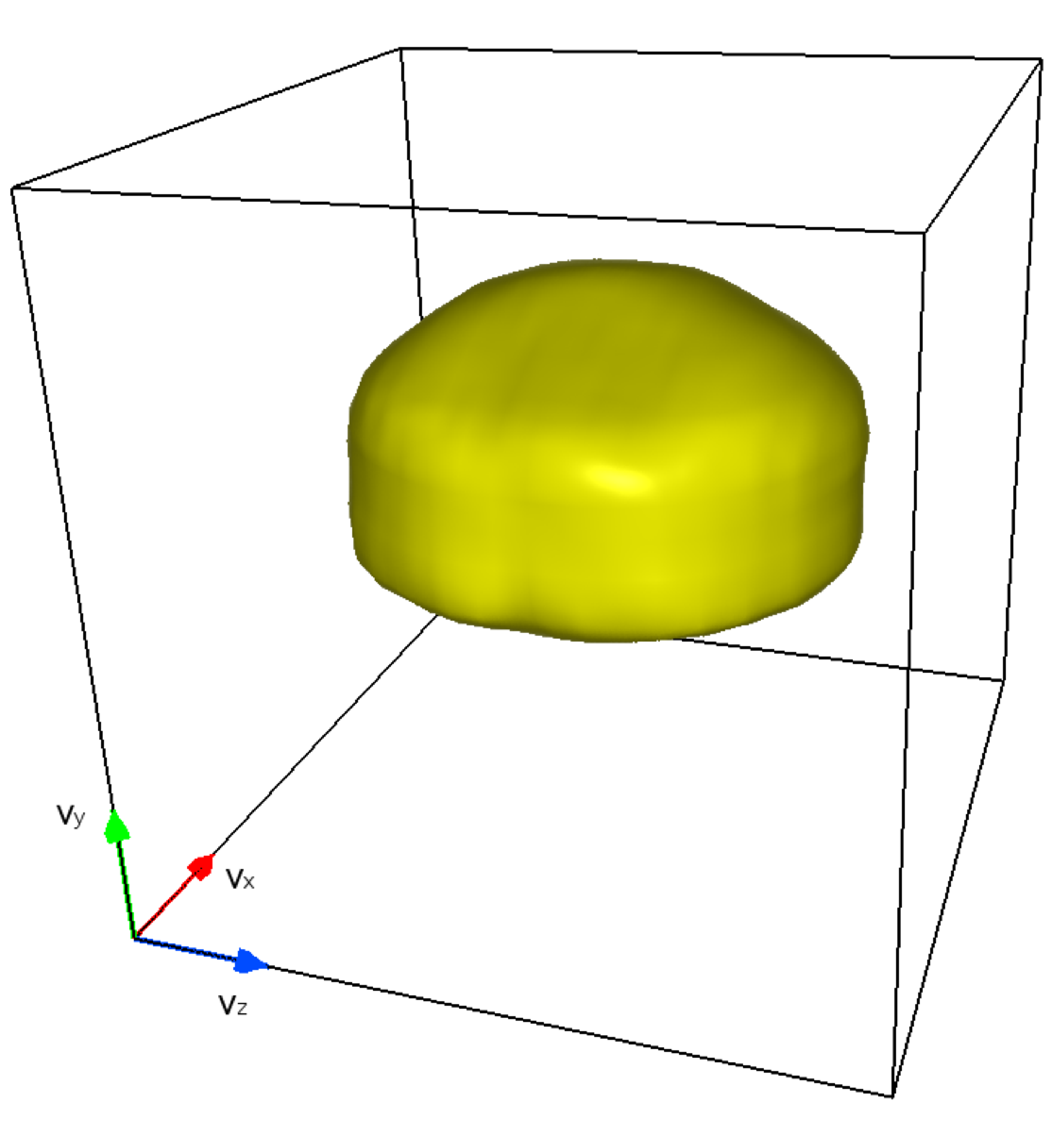} &
\epsfxsize=7.cm \epsffile{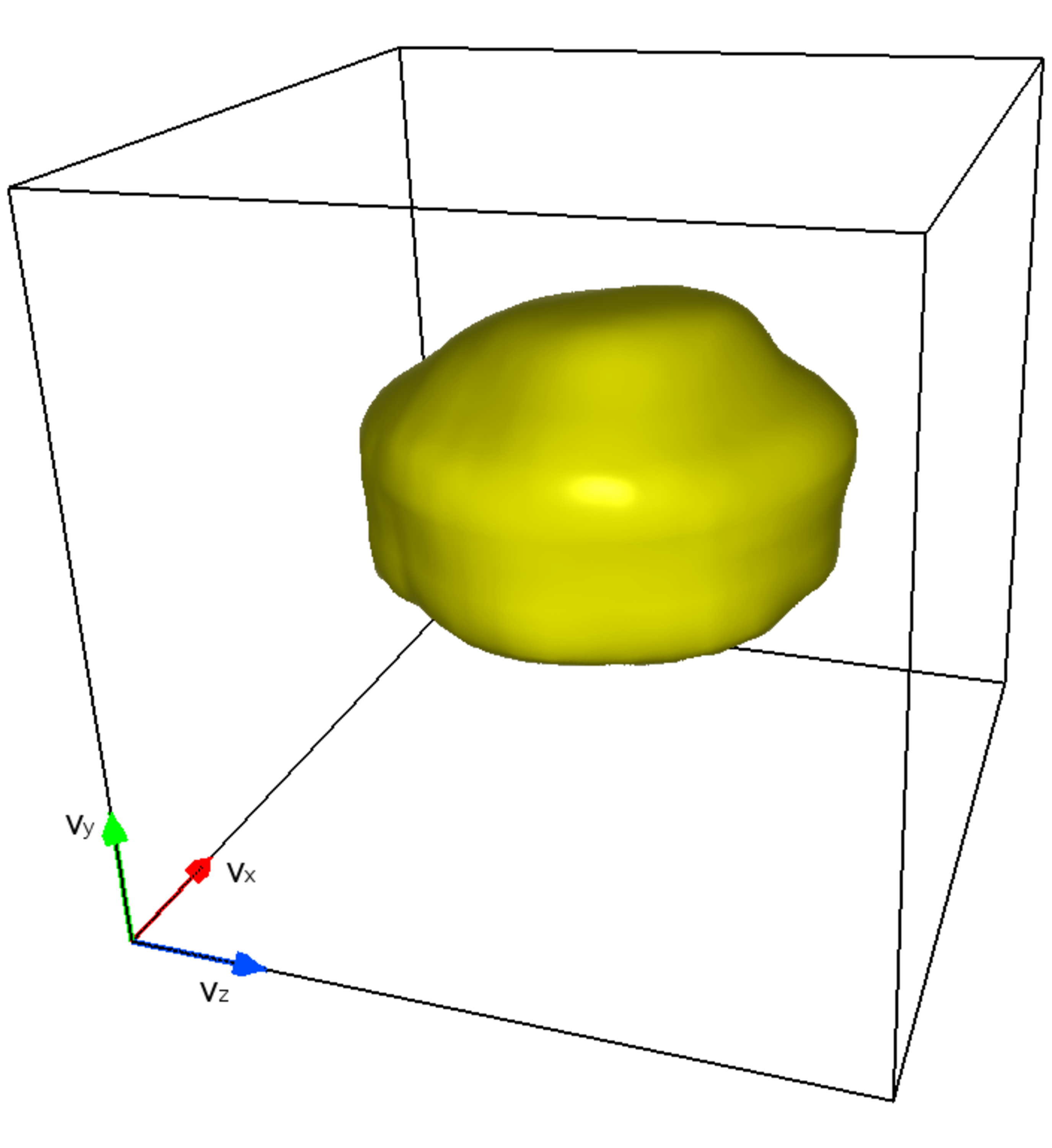} \\
\epsfxsize=7cm \epsffile{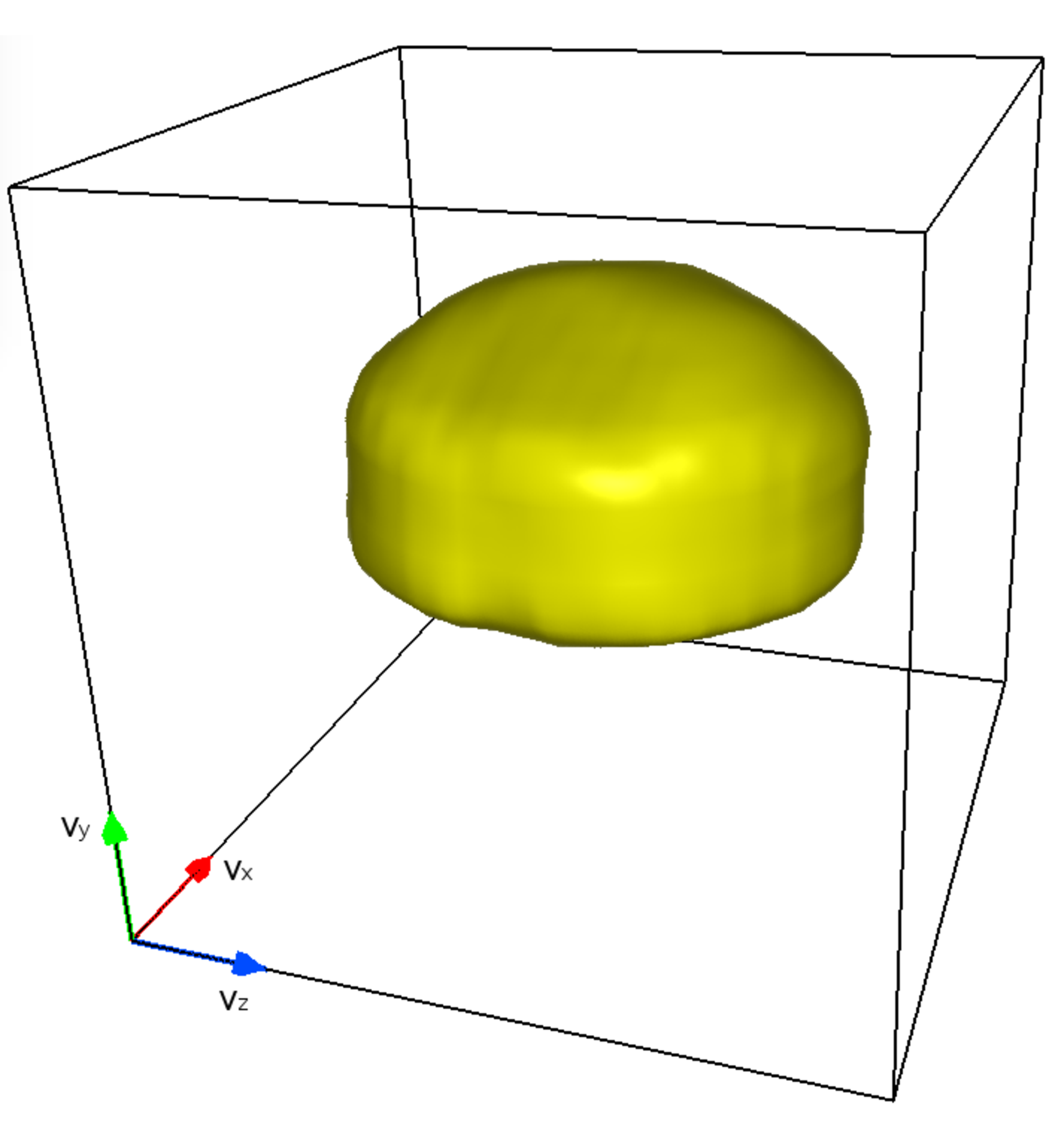} &
\epsfxsize=7cm \epsffile{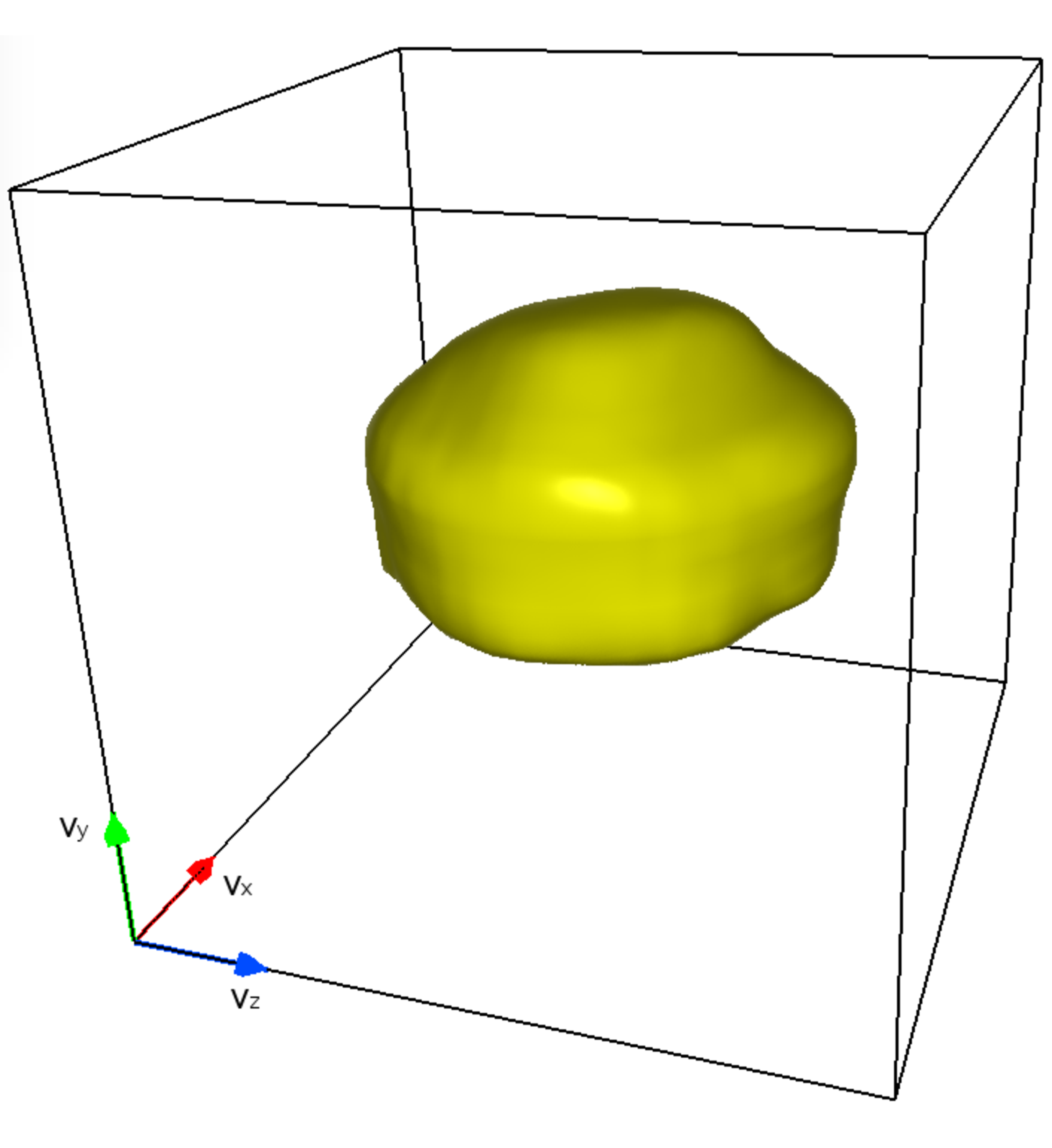} 
\end{array}$
\caption{(Color online) Thee-dimensional iso-surface plots of the proton velocity distribution at four different spatial positions $x_0\simeq 2.5, 3.5, 8.5, 10$ (top-left, top-right, bottom-left,
bottom-right,
respectively), for the nonlinear simulation with $a=0.25$, at $t=300$.}
\label{fig9} 
\end{figure*}

Finally, in Fig. \ref{fig9} we report the velocity iso-surfaces of the proton distribution function at $t=300$ evaluated at the spatial locations $x_0\simeq 2.5, 3.5, 8.5, 10$ (top-left,
top-right, bottom-left, bottom-right, respectively). In these three-dimensional plots it is clearly visible that the flattening along the $v_x$ direction, produced by nonlinear effects as discussed
above, makes the 3D proton velocity distributions look like flat disks (or pancakes), almost independently on the spatial location $x_0$ at which the 3D plot is considered. It is also worth noting
that the ring-like modulations of the 3D velocity distributions along the $v_y$ axis (especially visible in the top-right and bottom-right panels) presumably indicate $v_y$ resonance velocity shells.

\section{Summary and conclusions}
In this paper, we have analyzed numerically the kinetic features of the KAWs at large propagation angles, in typical conditions of the solar-wind environment, by employing the kinetic hybrid Vlasov-Maxwell code \cite{valentini07} in
1D-3V phase space configuration. Our kinetic simulations in nonlinear regime have been guided by a preliminary analysis of the two-fluid theory for the KAWs and by a set of linear simulations with
helped us to choose the simulation parameters, the initial condition, the numerical resolution and the form of the initial perturbations, in such a way to focus our study on the propagation of a
monochromatic KAW. 

While in linear regime the amplitude of the oscillations undergoes collisionless Landau damping, whose effect is larger for smaller propagation angles $\theta$, in the case of large initial amplitude
perturbations, the effects of damping is saturated. Then, the wave amplitude starts oscillating around an almost constant level with a period $\tau$, inversely proportional to the initial perturbation
amplitude. This phenomenology is clearly reminiscent of the nonlinear saturation of Landau damping in the electrostatic case, due to particle trapping \cite{oneil65}. In fact, also for the case of the
KAW it is possible to show that resonant protons can be trapped by a pseudo-potential and presumably trigger a physical process analogous to the trapping of particles in an electrostatic potential
well.

Thanks to the fact that the Eulerian HVM algorithm provides a clean noise-free description of the
phase space plasma dynamics, we pointed out how the resonant interaction between KAWs and protons can give rise to significant deformations of the proton distribution function, appearing as phase
space vortices and complicated structures. In particular, peculiar flat-top velocity profiles have been recovered in the velocity direction parallel to the wavevector. Moreover, the three-dimensional
iso-surface plots in velocity space have revealed that the proton velocity distribution assumes the typical shape of a flat disk, remarkably departing from the spherical isotropic Maxwellian configuration.

Finally, the numerical simulations presented in this paper suggest that when nonlinear processes of resonant wave-particle interaction are at play, describing the entire three-dimensional velocity
domain is crucial, since it allows the particle velocity distribution to freely model its shape, in response to its interaction with a large amplitude wave. The results discussed in this paper are
especially relevant in the field of space plasma physics, where the KAWs have recently gained an important role in the study of solar-wind turbulence dissipation and heating
\cite{bale05,sahraoui09,podesta12,salem12,chen13,kiyani13,howes08-1,schekochihin09,sahraoui12,gary04,howes08-2,tenbarge12}, at typical proton and electron kinetic scales. 

\section*{Acknowledgements}
The numerical simulations discussed in the present paper have been run on the Fermi supercomputer at CINECA (Bologna, Italy), within the ISCRA Class C project IsC15-KAWVLAS.  
F. V. and E. C. acknowledge the participation to the ISSI International Team 292 ``Kinetic Turbulence and Heating in the Solar Wind''.

\newpage   


\end{document}